\documentclass[aps,prl,twocolumn,superscriptaddress,floats,showpacs]{revtex4}

\usepackage{graphicx}
\usepackage{bm}
\usepackage{CJK}

\begin{document}

\title{Chiral Magnetic Effect and Chiral Phase Transition}

\author{Wei-jie Fu}
\email[]{wjfu@itp.ac.cn} \affiliation{Kavli Institute for
Theoretical Physics China (KITPC), Key Laboratory of Frontiers in
Theoretical Physics, Institute of Theoretical Physics, Chinese
Academy of Science, Beijing 100190, China}

\author{Yu-xin Liu}
\email[]{yxliu@pku.edu.cn} \affiliation{Department of Physics and
State Key Laboratory of Nuclear Physics and Technology, Peking
University, Beijing 100871, China} \affiliation{Center of
Theoretical Nuclear Physics, National Laboratory of Heavy Ion
Accelerator, Lanzhou 730000, China}

\author{Yue-liang Wu}
\email[]{ylwu@itp.ac.cn} \affiliation{Kavli Institute for
Theoretical Physics China (KITPC), Key Laboratory of Frontiers in
Theoretical Physics, Institute of Theoretical Physics, Chinese
Academy of Science, Beijing 100190, China}

\begin{abstract}
We study the influence of the chiral phase transition on the chiral
magnetic effect. The azimuthal charge-particle correlations as
functions of the temperature are calculated.  It is found that there
is a pronounced cusp in the correlations as the temperature reaches
its critical value for the QCD phase transition. It is predicted
that there will be a drastic suppression of the charge-particle
correlations as the collision energy in RHIC decreases to below a
critical value. We show then the azimuthal charge-particle
correlations can be the signal to identify the occurrence of the QCD
phase transitions in RHIC energy scan experiments.
\end{abstract}

\pacs{25.75.Nq  %Quark deconfinement, quark-gluon plasma production, and phase transitions
      11.30.Er, %Charge conjugation, parity, time reversal, and other discrete symmetries
      11.30.Rd, %Chiral symmetry
      11.30.Qc, %Spontaneous and radiative symmetry breaking
      }

\maketitle

The phase transitions of quantum chromodynamics (QCD), for example
the evolution between chiral symmetry breaking and its restoration,
the color deconfinement and confinement, have been one of the most
active topic in nuclear and particle physics in recent
years~\cite{LongRangePlan}. Such phase transitions can be driven by
the temperature and density of the system. It is then expected that
these phase transitions occur and the deconfined quark gluon phase
(QGP) is formed in ultrarelativistic heavy-ion
collisions~\cite{NPA757,Shuryak} (for example the current
experiments at the Relativistic Heavy Ion Collider (RHIC) and the
upcoming experiments at the Large Hadron Collider (LHC)) and in the
interior of neutron stars~\cite{Weber2005,Alford2008,Fu2008b}.
However, the explicit variation behavior of the signals to identify
the phase transitions with respect to the temperature and density
needs further investigations.

Recently, The STAR Collaboration at RHIC report their measurements
of azimuthal charged-particle correlations in Au + Au and Cu + Cu
collisions at $\sqrt{s_{NN}}=200\,\mathrm{GeV}$. They find a
significant signal consistent with the charge separation of quarks
along the system's orbital angular momentum
axis~\cite{Abelev2009a,Abelev2009b}. The observed charge separation
indicates that parity-odd domains, where the parity ($\mathcal{P}$)
symmetry is locally violated, might be created during the
relativistic heavy-ion
collisions~\cite{Kharzeev2006,Kharzeev2008,Fukushima2008}. The
charge separation is related with the so called ``chiral magnetic
effect'' which means that a magnetic field in the presence of
imbalanced chirality induces a current along the magnetic field, and
therefore results in that positive charge is separated from negative
charge along the magnetic field~\cite{Fukushima2008}.

Now that the chiral magnetic effect can be observed through the
measurements of azimuthal charged-particle correlations in the
relativistic heavy-ion collisions, a natural question arises, i.e.
whether can we detect the properties of the QCD phase transitions,
especially the chiral phase transition through the observations of
the chiral magnetic effect? To answer this question, we have to
study how the chiral magnetic effect or the charge separation effect
is influenced by the chiral phase transition. This is our central
subject in this letter.

In this work, we will study the chiral magnetic effect and the QCD
phase transitions in the 2+1 flavor Polyakov--Nambu--Jona-Lasinio
(PNJL) model~\cite{Fu2008}. The validity of the PNJL model has been
confirmed in a series of works by confronting the PNJL results with
the lattice QCD data~\cite{Fu2008,Weise20067,Fu2009}. The PNJL model
not only has the chiral symmetry and the dynamical breaking
mechanism of this symmetry, which are same as the conventional
Nambu--Jona-Lasinio model, but also include the effect of color
confinement through the Polyakov loop. Therefore, the PNJL model is
very appropriate to describe the QCD phase transitions at finite
temperature and/or density.

The Lagrangian density for the 2+1 flavor PNJL model is given as
\begin{eqnarray}
\mathcal{L}_{\mathrm{PNJL}}&=&\bar{\psi}(i\gamma_{\mu}D^{\mu}-\hat{m}_{0})\psi
 +G\sum_{a=0}^{8}\Big[(\bar{\psi}\tau_{a}\psi)^{2}\nonumber \\
 &&+(\bar{\psi}i\gamma_{5}\tau_{a}\psi)^{2}\Big]
-K\Big[\textrm{det}_{f}\big(\bar{\psi}(1+\gamma_{5})\psi\big)\nonumber \\
 &&+\textrm{det}_{f}\big(\bar{\psi}(1-\gamma_{5})\psi\big)\Big]
 -\mathcal{U}(\Phi,\Phi^{*} \, ,T),\label{lagragian}
\end{eqnarray}
where $\psi=(\psi_{u},\psi_{d},\psi_{s})^{T}$ is the three-flavor
quark field, $D^{\mu}=\partial^{\mu}-iA^{\mu}$ with $\quad
A^{\mu}=\delta^{\mu}_{0}A^{0}$,
$A^{0}=g\mathcal{A}^{0}_{a}\frac{\lambda_{a}}{2}=-iA_{4}$.
$\lambda_{a}$ are the Gell-Mann matrices in color space;
$\hat{m}_{0}=\textrm{diag}(m_{0}^{u},m_{0}^{d},m_{0}^{s})$ is the
three-flavor current quark mass matrix. In this work, we take
$m_{0}^{u}=m_{0}^{d}\equiv m_{0}^{l}$, while keep $m_{0}^{s}$ being
larger than $m_{0}^{l}$. $\mathcal{U}(\Phi,\Phi^{*},T)$ in the PNJL
Lagrangian density is the Polyakov-loop effective potential, which
is expressed in terms of the traced Polyakov-loop
$\Phi=(\mathrm{Tr}_{c}L)/N_{c}$ and its conjugate
$\Phi^{*}=(\mathrm{Tr}_{c}L^{\dag})/N_{c}$. In this work, we use the
Polyakov-loop effective potential which is a polynomial in $\Phi$
and $\Phi^{*}$~\cite{Weise20067}, given by
\begin{eqnarray}
\!\!\!\!\frac{\mathcal{U}(\Phi,\Phi^{*},T)}{T^{4}}\!\!&=&\!\!
-\!\frac{b_{2}(T)}{2}\Phi^{*}\Phi\! -\!\!\frac{b_{3}}{6}
(\Phi^{3}\!\!+\!\!{\Phi^{*}}^{3})+\!\!\frac{b_{4}}{4}(\Phi^{*}\Phi)^{2}\!,
\end{eqnarray}
with
\begin{equation}
b_{2}(T)=a_{0}+a_{1}\left(\frac{T_{0}}{T}\right)+a_{2}
{\left(\frac{T_{0}}{T}\right)}^{2}
+a_{3}{\left(\frac{T_{0}}{T}\right)}^{3}.
\end{equation}
Parameters in the effective potential are fixed by fitting the
thermodynamical behavior of the pure-gauge QCD obtained from the
lattice simulations. Their values are $a_{0}=6.75$, $a_{1}=-1.95$,
$a_{2}=2.625$, $a_{3}=-7.44$, $b_{3}=0.75$ and $b_{4}=7.5$. The
parameter $T_{0}$ is the critical temperature for the deconfinement
phase transition to take place in the pure-gauge QCD and $T_{0}$ is
chosen to be $270\,\mathrm{MeV}$ according to the lattice
calculations. Furthermore, we also need to determine the five
parameters in the quark sector of the model, which are
$m_{0}^{l}=5.5\;\mathrm{MeV}$, $m_{0}^{s}=140.7\;\mathrm{MeV}$,
$G\Lambda^{2}=1.835$, $K\Lambda^{5}=12.36$ and
$\Lambda=602.3\;\mathrm{MeV}$. They are fixed by fitting
$m_{\pi}=135.0\;\mathrm{MeV}$, $m_{K}=497.7\;\mathrm{MeV}$,
$m_{\eta^{\prime}}=957.8\;\mathrm{MeV}$ and
$f_{\pi}=92.4\;\mathrm{MeV}$~\cite{Rehberg1996}.

In the parity-odd domains which are created during relativistic
heavy-ion collisions, the number of left- and right-hand quarks is
different because of the axial anomaly. In this work we introduce
the chiral chemical potential $\mu_{5}$ to study the left-right
asymmetry following the method of Ref.~\cite{Fukushima2008}, where
the chiral chemical potential $\mu_{5}$ is related with the
effective theta angle of the $\theta$-vacuum through
$\mu_{5}=\partial_{0}\theta/2N_{f}$ and $N_{f}$ is the number of
flavor. Consequently, we should add the following term
\begin{equation}
\bar{\psi}\hat{\mu}_{5}\gamma^{0}\gamma^{5}\psi
\end{equation}
to the Lagrangian density in Eq.~(\ref{lagragian}), where
$\hat{\mu}_{5}=\textrm{diag}(\mu_{5}^{u},\mu_{5}^{d},\mu_{5}^{s})$.
Next, we consider the case that a homogenous magnetic field $B$ is
along the direction of the orbital angular momentum of the system
produced in a non-central heavy-ion collision. In the following we
denote this direction with $z$-direction and particle momentum in
this direction with $p_{3}$.

In the mean field approximation, the thermodynamical potential
density for the 2+1 flavor quark system under a homogeneous
background magnetic field $B$ and with left-right asymmetry is given
by
\begin{eqnarray}
\Omega &=&-N_{c}\sum_{f=u,d,s}\frac{|q_{f}|e
B}{2\pi}\sum_{n=0}^{\infty}\sum_{s=\pm 1}\int\frac{d
p_{3}}{2\pi}\bigg(E_{f}\nonumber \\
&&+\frac{T}{3}\ln\Big\{1+3\Phi^{*}\exp\big[-\big(E_{f}-\mu_{f}-\frac{s|\epsilon_{f}|}{E_{f}}\mu_{5}^{f}\big)/T\big]\nonumber \\
&&+3\Phi
\exp\big[-2\big(E_{f}-\mu_{f}-\frac{s|\epsilon_{f}|}{E_{f}}\mu_{5}^{f}\big)/T\big]\nonumber\\
&&+\exp\big[-3\big(E_{f}-\mu_{f}-\frac{s|\epsilon_{f}|}{E_{f}}\mu_{5}^{f}\big)/T\big]\Big\}+\frac{T}{3}\ln\Big\{1\nonumber\\
&&+3\Phi
\exp\big[-\big(E_{f}+\mu_{f}-\frac{s|\epsilon_{f}|}{E_{f}}\mu_{5}^{f}\big)/T\big]\nonumber\\
&&+3\Phi^{*}\exp\big[-2\big(E_{f}+\mu_{f}-\frac{s|\epsilon_{f}|}{E_{f}}\mu_{5}^{f}\big)/T\big]\nonumber\\
&&+\exp\big[-3\big(E_{f}+\mu_{f}-\frac{s|\epsilon_{f}|}{E_{f}}\mu_{5}^{f}\big)/T\big]\Big\}\bigg)\nonumber\\
&&+2G({\phi_{u}}^{2}
+{\phi_{d}}^{2}+{\phi_{s}}^{2})-4K\phi_{u}\,\phi_{d}\,\phi_{s}\nonumber\\
&&+\mathcal{U}(\Phi,\Phi^{*},T),\label{thermopotential}
\end{eqnarray}
where
\begin{equation}
|\epsilon_{f}|=\sqrt{2n|q_{f}|e B+p_{3}^{2}}, \label{momentum}
\end{equation}
\begin{equation}
E_{f}=\sqrt{2n|q_{f}|e B+p_{3}^{2}+M_{f}^{2}}.\label{energy}
\end{equation}
with $q_{i}(i=u,d,s)$ being the electric charge in unit of
elementary charge $e$ for the quark of flavor $i$ and the
constituent mass $M_{i}$ reading
\begin{equation}
M_{i}=m_{0}^{i}-4G\langle\bar{\psi}\psi\rangle_{i}+2K\langle\bar{\psi}\psi\rangle_{j}\,\langle\bar{\psi}\psi\rangle_{k},\label{constituentmass}
\end{equation}
and $\langle\bar{\psi}\psi\rangle_{i}$ is the chiral condensate. In
Eq.~(\ref{thermopotential}) we also include the quark chemical
potential $\mu_{i}$. The momenta of charged particles in the
longitudinal direction, i.e., the $z$-direction, are not influenced
by the background magnetic field and $p_{3}$ in the expression of
the thermodynamical potential density in Eq.~(\ref{thermopotential})
is continuous; while the momenta in the transverse plane are
discretized due to the magnetic field effect. $|\epsilon_{f}|$ in
Eq.~(\ref{momentum}) is similar to the magnitude of the momentum in
free space. $s$ (for fermion and for anti-fermion is $-s$) in
Eq.~(\ref{thermopotential}) is the helicity of particle and we
should emphasize that at the lowest order of the transverse quantum
number, i.e., $n=0$, the quark spin only has one value in the
$z$-direction, which means that charged particles in the lowest
transverse level are polarized by the external magnetic field;
however particles in higher levels, i.e., $n>0$, are not polarized.
Therefore, the charge separation effect only comes from quarks in
the lowest transverse level.

In order to relate our calculations with observable in heavy-ion
collisions, we define $\Delta_{+}$ ($\Delta_{-}$) to be the positive
(negative) charge difference in unit of $e$ ($-e$) between on each
side of the $z=0$ plane, which is also the reaction plane. Here we
use the notations in Ref~\cite{Kharzeev2008}. Taking particles with
positive elementary electric charge $e$ for example, we can express
$\Delta_{+}$ as
\begin{eqnarray}
\Delta_{+}&=&\int d^{3}x
(\bar{\psi}\gamma^{0}\gamma^{5}\psi)_{n=0}\nonumber\\
&=&\int
d^{3}x({\bar{\psi}}_{R}\gamma^{0}\psi_{R}-{\bar{\psi}}_{L}\gamma^{0}\psi_{L})_{n=0},\label{delta}
\end{eqnarray}
where
\begin{equation}
\psi_{R}=\frac{1+\gamma^{5}}{2}\psi \quad \mathrm{and}\quad
\psi_{L}=\frac{1-\gamma^{5}}{2}\psi .
\end{equation}
The subscript $n=0$ in Eq.~(\ref{delta}) indicates that only the
lowest transverse level states contribute to the $\Delta_{+}$. In
the same way, we can obtain $\Delta_{\pm}$ for the 2+1 quark system.
We take $\Delta_{+}$ for example once more, which is given as
\begin{eqnarray}
\Delta_{+}&=&VN_{c}\frac{e
B}{4\pi^{2}}\bigg\{q_{u}^{2}\int_{0}^{\infty}d
p_{3}\frac{p_{3}}{E_{u}}\Big[f\big(E_{u}-\mu_{u}\nonumber\\
&&-\frac{p_{3}}{E_{u}}\mu_{5}^{u}\big)-f\big(E_{u}-\mu_{u}+\frac{p_{3}}{E_{u}}\mu_{5}^{u}\big)\Big]\nonumber\\
&&+q_{d}^{2}\int_{0}^{\infty}d
p_{3}\frac{p_{3}}{E_{d}}\Big[\bar{f}\big(E_{d}+\mu_{d}-\frac{p_{3}}{E_{d}}\mu_{5}^{d}\big)\nonumber\\
&&-\bar{f}\big(E_{d}+\mu_{d}+\frac{p_{3}}{E_{d}}\mu_{5}^{d}\big)\Big]\nonumber\\&&
+q_{s}^{2}\int_{0}^{\infty}d
p_{3}\frac{p_{3}}{E_{s}}\Big[\bar{f}\big(E_{s}+\mu_{s}-\frac{p_{3}}{E_{s}}\mu_{5}^{s}\big)\nonumber\\
&&-\bar{f}\big(E_{s}+\mu_{s}+\frac{p_{3}}{E_{s}}\mu_{5}^{s}\big)\Big]\bigg\},\label{delta1}
\end{eqnarray}
where $V$ is the volume of the system, $E_{f}$ is given by
Eq.~(\ref{energy}) with $n=0$, and
\begin{equation}
f(x)=\frac{\Phi^{*}e^{-x/T}+2\Phi
e^{-2x/T}+e^{-3x/T}}{1+3\Phi^{*}e^{-x/T}+3\Phi e^{-2x/T}+e^{-3x/T}}
\end{equation}
and
\begin{equation}
\bar{f}(x)=\frac{\Phi e^{-x/T}+2\Phi^{*}
e^{-2x/T}+e^{-3x/T}}{1+3\Phi e^{-x/T}+3\Phi^{*}
e^{-2x/T}+e^{-3x/T}}.
\end{equation}
In fact, we can also obtain Eq.~(\ref{delta1}) through
differentiating the thermodynamical potential in
Eq.~(\ref{thermopotential}) with respect to the chiral chemical
potential $\mu_{5}^{i}$ and summing over the contributions from
positive quarks or anti-quarks.

In the same way, differentiating the thermodynamical potential with
respect to the quark chemical potential $\mu_{i}$ and summing over
contributions from the three-flavor positive quarks or anti-quarks
we obtain the total positive electric charge number $N_{+}$ in unit
of $e$, i.e.,
\begin{eqnarray}
N_{+}\!\!&=&\!\!VN_{c}\frac{e B}{2\pi}\sum_{n=0}^{\infty}\sum_{s=\pm
1}\Big[q_{u}^{2}\int\frac{d
p_{3}}{2\pi}f\big(E_{u}-\mu_{u}-\frac{s|\epsilon_{u}|}{E_{u}}\mu_{5}^{u}\big)\nonumber\\
&&+q_{d}^{2}\int\frac{d
p_{3}}{2\pi}\bar{f}\big(E_{d}+\mu_{d}-\frac{s|\epsilon_{d}|}{E_{d}}\mu_{5}^{d}\big)\nonumber\\
&&+q_{s}^{2}\int\frac{d p_{3}}{2\pi}\bar{f}\big(E_{s}+\mu_{s}
-\frac{s|\epsilon_{s}|}{E_{s}}\mu_{5}^{s}\big)\Big].
\end{eqnarray}
Similarly, the total negative electric charge number $N_{-}$ in unit
of $-e$ can also be obtained.

In the experiments of heavy-ion collisions, the azimuthal
charged-particle correlations, i.e., $\langle
\cos(\phi_{\alpha}+\phi_{\beta}-2\Psi_{RP})\rangle$, are used to
detect the $\mathcal{P}$-violating
effect~\cite{Voloshin2004,Abelev2009a,Abelev2009b}. Here $\phi$ and
$\Psi_{RP}$ are the azimuthal angles of the particles and reaction
plane, respectively. $\alpha$, $\beta$ represent electric charge $+$
or $-$. With the notation $a_{\alpha\beta}\equiv-\langle
\cos(\phi_{\alpha}+\phi_{\beta}-2\Psi_{RP})\rangle$, it can be shown
that~\cite{Kharzeev2008}
\begin{equation}
a_{++}=\frac{\pi^{2}}{16}\frac{\langle\Delta_{+}^{2}\rangle}{N_{+}^{2}},
\quad
a_{--}=\frac{\pi^{2}}{16}\frac{\langle\Delta_{-}^{2}\rangle}{N_{-}^{2}},\label{a1}
\end{equation}
and
\begin{equation}
a_{+-}=\frac{\pi^{2}}{16}\frac{\langle\Delta_{+}\Delta_{-}\rangle}{N_{+}N_{-}},\label{a2}
\end{equation}
where the azimuthal angle distribution of the charged particles is
assumed to be
\begin{equation}
\frac{d N_{\pm}}{d
\phi}=\frac{1}{2\pi}N_{\pm}+\frac{1}{4}\Delta_{\pm}\sin\phi.
\end{equation}
Since we mainly focus on the influence of the QCD phase transitions,
especially the chiral phase transition, on the chiral magnetic
effect in this work, we will neglect the screening suppression
effect due to the final state interactions~\cite{Kharzeev2008} and
make $\mu_{i}=0$, then we have $a_{++}=a_{--}=-a_{+-}$. Therefore,
we only study $a_{++}$ in the following.

%Since we mainly focus on the influence of the QCD phase transitions,
%especially the chiral phase transition, on the chiral magnetic
%effect in this work, we just replace the averaged
%$\langle\Delta_{+}^{2}\rangle$, $\langle\Delta_{-}^{2}\rangle$, and
%$\langle\Delta_{+}\Delta_{-}\rangle$ in Eqs.~(\ref{a1})~(\ref{a2})
%with $\Delta_{+}^{2}$, $\Delta_{-}^{2}$, and $\Delta_{+}\Delta_{-}$
%at finit respectively .

Minimizing the thermodynamical potential in
Eq.~(\ref{thermopotential}) with respect to three-flavor quark
condensates, $\Phi$, and $\Phi^{*}$, we obtain a set of equations of
motion. We neglect the influence of the magnetic field on these
equations of motion in our numerical calculations, since the
magnetic field ($eB=10^{2}\sim 10^{4}\,\mathrm{MeV}^{2}$ in the
non-central heavy-ion collisions~\cite{Kharzeev2008}) has little
impact on these equations of motion.

\begin{figure}[!htb]
\includegraphics[scale=0.6]{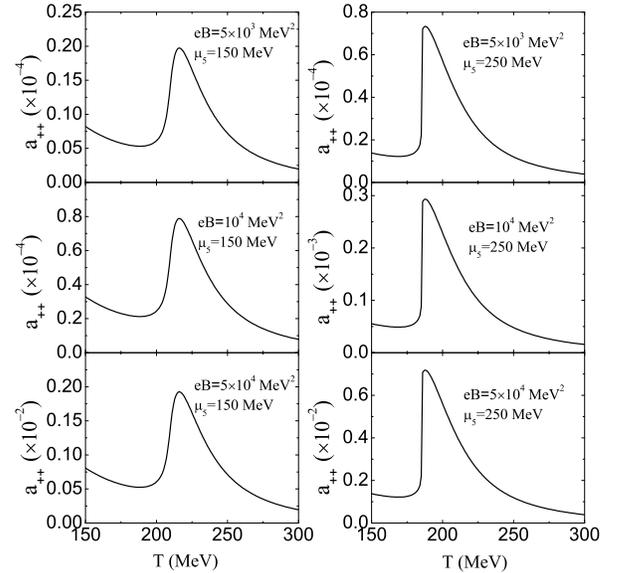}
\caption{Correlation $a_{++}$ as function of the temperature
calculated in the PNJL model with $\mu_{5}=150\,\mathrm{MeV}$ (left
panel) and $\mu_{5}=250\,\mathrm{MeV}$ (right panel). The magnetic
field corresponds to $e B=5\times 10^{3}$, $10^{4}$, and $5\times
10^{4}\,\mathrm{MeV}^{2}$ from top to bottom,
respectively.}\label{f1}
\end{figure}

In Fig.~\ref{f1} we show $a_{++}$ defined in Eq.~(\ref{a1}) as
function of the temperature at several values of the chiral chemical
potential $\mu_{5}$ (here
$\mu_{5}\equiv\mu_{5}^{u}=\mu_{5}^{d}=\mu_{5}^{s}$) and the magnetic
field strength. We find that there is a pronounced cusp in $a_{++}$
at the critical temperature during the chiral phase transition (the
critical temperature $T_{c}=209\,\mathrm{MeV}$ for
$\mu_{5}=150\,\mathrm{MeV}$ and $T_{c}=185\,\mathrm{MeV}$ for
$\mu_{5}=250\,\mathrm{MeV}$ in the PNJL model). From the
Fig.~\ref{f1} one can also find that although the value of $a_{++}$
is proportional to the square of the magnetic field strength, the
shape of the curve for $a_{++}$ as function of temperature is almost
independent of the magnetic field strength. Furthermore, the cusp at
the critical temperature in the curve becomes much sharper with the
increase of the chiral chemical potential. With the decrease of the
temperature, when the temperature is below $T_{c}$, chiral symmetry
is dynamically broken and quarks obtain masses. Since the axial
anomaly can be suppressed by the mass effect, which has been
discussed in detail in Ref.~\cite{Ma2006}, the chiral magnetic
effect can also be suppressed by large constituent quark masses.
Therefore, the azimuthal charged-particle correlations described by
$a_{++}$ ($a_{--}$ and $a_{+-}$) defined in
Eqs.~(\ref{a1})~(\ref{a2}) are quite decreased once the temperature
is below the critical temperature. It can been seen from
Fig.~\ref{f1} that, when the temperature is above $T_{c}$, $a_{++}$
decreases with the increase of the temperature, which is because
higher temperature makes it more difficult to polarize quarks with
magnetic field and thus suppresses the charge separation effect.

What do our calculated results imply in future energy scanning
experiments of heavy-ion collisions? With the decrease of the
heavy-ion collision energy, the temperature of the QGP produced in
the fireball at early stage is also decreased. Since the magnetic
field produced in non-central collisions decays rapidly with
time~\cite{Kharzeev2008}, the observed charge separation mainly
carries the information of the QGP at early stage. Therefore, we
expect that the azimuthal charged-particle correlations (especially
for the same charge correlations, because the opposite charge
correlations are suppressed by final state interactions) increase as
the collision energy is lowered. However, when the collision energy
is lowered to the value that cannot drive the chiral phase
transition, it is expected that the azimuthal charged-particle
correlations are quite suppressed. Therefore, we can employ the
charge separation effect to locate where the QCD phase transitions
occur.

In summary, we have studied the influence of the QCD phase
transitions on the chiral magnetic effect. The azimuthal
charge-particle correlations as functions of the temperature are
calculated in the PNJL model. It is found that there is a pronounced
cusp in the azimuthal charge-particle correlations around the
critical temperature of the chiral phase transition. We predict that
there will be a sudden suppression of the charge-particle
correlations with the decrease of the collision energy. It indicates
that azimuthal charge-particle correlations can be a signal to
identify chiral phase transition in the energy scan experiment in
RHIC.

%\bigskip

\begin{acknowledgments}
This work was supported by the National Natural Science Foundation
of China under contract Nos. 10425521, 10675007, 10935001, the Major
State Basic Research Development Program under contract Nos.
G2007CB815000. One of the authors (W.J.F.) would also acknowledge
the financial support from China Postdoctoral Science Foundation No.
20090460534.
\end{acknowledgments}


\begin{thebibliography}{50}
\bibitem{LongRangePlan}
DOE/NSF Nuclear Science Advisory Committee, arXiv:0809.3137.

\bibitem{NPA757}
I.~Arsene \textit{et al}, Nucl. Phys. {\bf A 757}, 1 (2005);
%
%\bibitem{Back2005}
B.~B.~Back \textit{et al}, Nucl. Phys. {\bf A 757}, 28 (2005);
%
%\bibitem{Adams2005}
J.~Adams \textit{et al}, Nucl. Phys. {\bf A 757}, 102 (2005);
%
%\bibitem{Adcox2005}
K.~Adcox \textit{et al}, Nucl. Phys. {\bf A 757}, 184 (2005).

\bibitem{Shuryak}
  E. V. Shuryak,
     Prog. Part. Nucl. Phys. {\bf 53}, 273 (2004);
  M. Gyulassy, and L. McLerran,
     Nucl. Phys. A {\bf 750}, 30 (2005);
  E. Shuryak, Nucl. Phys. A {\bf 750}, 64 (2005).

\bibitem{Weber2005} F. Weber,
Prog. Part. Nucl. Phys. {\bf 54}, 193 (2005).

\bibitem{Alford2008}
M.~Alford, A.~Schmitt, K.~Rajagopal, and T. Sch\"{a}fer, Rev. Mod.
Phys. {\bf 80}, 1455 (2008).

\bibitem{Fu2008b}
W.~J.~Fu, H.~Q.~Wei, and Y.~X.~Liu, Phys. Rev. Lett. {\bf 101},
181102 (2008).

\bibitem{Abelev2009a}
B.~I.~Abelev \textit{et al}, Phys. Rev. Lett. {\bf 103}, 251601
(2009).

\bibitem{Abelev2009b}
B.~I.~Abelev \textit{et al}, arXiv:0909.1717 [nucl-ex].

\bibitem{Kharzeev2006}
D.~Kharzeev, Phys. Lett. {\bf B 633}, 260 (2006).

\bibitem{Kharzeev2008}
D.~E.~Kharzeev, L.~D.~McLerran, and H.~J.~Warringa, Nucl. Phys. {\bf
A 803}, 227 (2008).

\bibitem{Fukushima2008}
K.~Fukushima, D.~E.~Kharzeev, and H.~J.~Warringa, Phys. Rev. {\bf D
78}, 074033 (2008).

\bibitem{Fu2008}
W.~J.~Fu, Z.~Zhang, and Y.~X.~Liu, Phys. Rev. {\bf D 77}, 014006
(2008); K.~Fukushima, Phys. Rev. {\bf D 77}, 114028, (2008).


\bibitem{Weise20067}
C.~Ratti, M.~A.~Thaler, and W.~Weise, Phys. Rev. D {\bf 73}, 014019
(2006);
%
%\bibitem{Roner2007}
S.~R\"{o}{\ss}ner, C.~Ratti, and W.~Weise, Phys. Rev. D {\bf 75},
034007 (2007);
%
%\bibitem{Ghosh2006}
%
S.~K.~Ghosh, T.~K.~Mukherjee, M.~G.~Mustafa, and R.~Ray, Phys. Rev.
D {\bf 73}, 114007 (2006); S.~Mukherjee, M.~G.~Mustafa, and R.~Ray,
Phys. Rev. D {\bf 75}, 094015 (2007).


\bibitem{Fu2009}
W.~J.~Fu, Y.~X.~Liu, and Y.~L.~Wu, Phys. Rev. {\bf D 81}, 014028
(2010).

\bibitem{Rehberg1996}
P.~Rehberg, S.~P.~Klevansky, and J.~H\"{u}fner, Phys. Rev. {\bf C
53}, 410 (1996).

\bibitem{Voloshin2004}
S.~A.~Voloshin, Phys. Rev. {\bf C 70}, 057901 (2004).

\bibitem{Ma2006}
Y.~L.~Ma, and Y.~L.~Wu, Int. J. Mod. Phys. {\bf A 21}, 6383 (2006).

\end{thebibliography}
\end{document}